\documentclass[prl,twocolumn,superscriptaddress,nobibnotes,letterpaper]{revtex4}

\usepackage[pdftex]{graphicx}
\usepackage[pdftex]{epsfig}
\usepackage{amsmath}
\usepackage{amssymb}
\usepackage{amsfonts}
\usepackage{color}
\usepackage{wrapfig}
\usepackage{eucal}
\usepackage{hhline}
\usepackage{threeparttable}
\usepackage{supertabular}
\usepackage{multirow}
\usepackage{tabularx}

\begin{document}
	
\pagenumbering{arabic}

\title{Hanbury Brown and Twiss interferometry of single phonons\\
from an optomechanical resonator}\thanks{This work was published in Science \textbf{358}, 203--206 (2017).}

\author{Sungkun Hong}\thanks{These authors contributed equally to this work.}
\affiliation{Vienna Center for Quantum Science and Technology (VCQ), Faculty of Physics, University of Vienna, A-1090 Vienna, Austria}
\author{Ralf Riedinger}\thanks{These authors contributed equally to this work.}
\affiliation{Vienna Center for Quantum Science and Technology (VCQ), Faculty of Physics, University of Vienna, A-1090 Vienna, Austria}
\author{Igor Marinkovi\'{c}}\thanks{These authors contributed equally to this work.}
\affiliation{Kavli Institute of Nanoscience, Delft University of Technology, 2628CJ Delft, Netherlands}
\author{Andreas Wallucks}\thanks{These authors contributed equally to this work.}
\affiliation{Kavli Institute of Nanoscience, Delft University of Technology, 2628CJ Delft, Netherlands}
\author{Sebastian G. Hofer}
\affiliation{Vienna Center for Quantum Science and Technology (VCQ), Faculty of Physics, University of Vienna, A-1090 Vienna, Austria}
\author{Richard A.\ Norte}
\affiliation{Kavli Institute of Nanoscience, Delft University of Technology, 2628CJ Delft, Netherlands}
\author{Markus Aspelmeyer}
\email{markus.aspelmeyer@univie.ac.at}
\affiliation{Vienna Center for Quantum Science and Technology (VCQ), Faculty of Physics, University of Vienna, A-1090 Vienna, Austria}
\author{Simon Gr\"oblacher}
\email{s.groeblacher@tudelft.nl}
\affiliation{Kavli Institute of Nanoscience, Delft University of Technology, 2628CJ Delft, Netherlands}

\begin{abstract}
Nano- and micromechanical solid-state quantum devices have become a focus of attention. Reliably generating nonclassical states of their motion is of interest both for addressing fundamental questions about macroscopic quantum phenomena and for developing quantum technologies in the domains of sensing and transduction. We used quantum optical control techniques to conditionally generate single-phonon Fock states of a nanomechanical resonator. We performed a Hanbury Brown and Twiss--type experiment that verified the nonclassical nature of the phonon state without requiring full state reconstruction. Our result establishes purely optical quantum control of a mechanical oscillator at the single-phonon level.
\end{abstract}

\maketitle

Intensity correlations in electromagnetic fields have been pivotal in the development of modern quantum optics. The experiments by Hanbury Brown and Twiss were a particular milestone that connected the temporal and spatial coherence properties of a light source with the second-order intensity autocorrelation function $g^{(2)}(\tau, x)$~\cite{HanburyBrown1956a,HanburyBrown1956,Morgan1966}. In essence, $g^{(2)}$ correlates intensities measured at times differing by $\tau$ or at locations differing by $x$ and hence is a measure of their joint detection probability. At the same time, these correlations allow the quantum nature of the underlying field to be inferred directly. For example, a classical light source of finite coherence time can only exhibit positive correlations at a delay of $\tau\approx0$ in the joint intensity detection probability, leading to bunching in the photon arrival time. This result holds true for all bosonic fields. Fermions, on the other hand, exhibit negative correlations and hence antibunching in the detection events~\cite{Henny1999,Oliver1999,Jeltes2007}, which is a manifestation of the Pauli exclusion principle. A bosonic system needs to be in a genuine nonclassical state to exhibit antibunching. The canonical example is a single-photon (Fock) state, for which $g^{(2)}(\tau=0)=0$ because no joint detection can take place~\cite{Grangier1986}. For this reason, measuring $g^{(2)}$ has become a standard method to characterize the purity of single-photon sources~\cite{Eisamana2011}. In general, $g^{(2)}(\tau)$ carries a wealth of information on the statistical properties of a bosonic field with no classical analogue~\cite{Zou1990,Davidovich1996} -- specifically sub-Poissonian counting statistics [$g^{(2)}(0)<1$] and antibunching [$g^{(2)}(\tau) \geq g^{(2)}(0)$] -- all of which have been demonstrated successfully with quantum states of light~\cite{Kimble1977,Short1983}.

\begin{figure*}[t]
	\begin{center}
		\includegraphics[width=1.7\columnwidth]{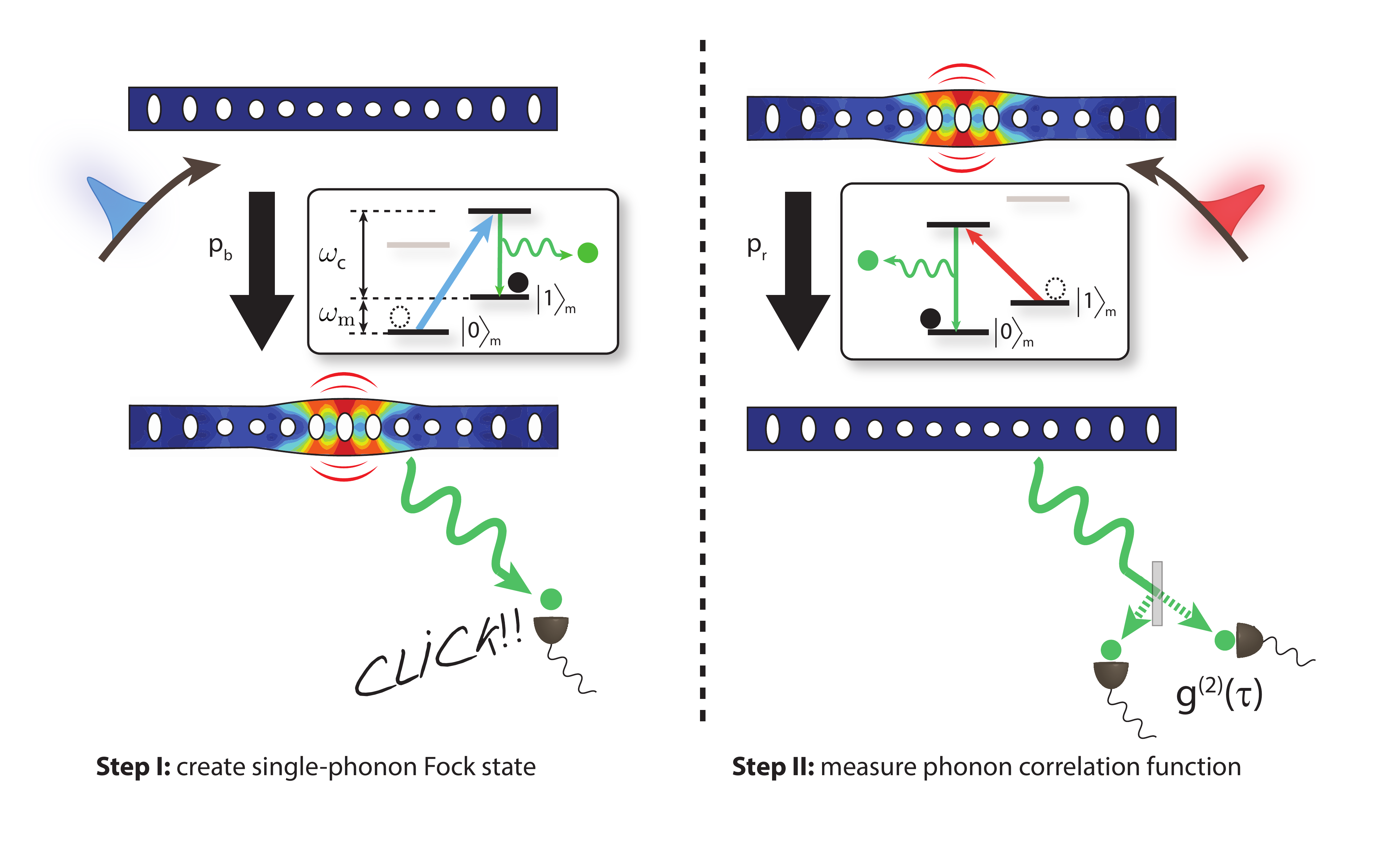}
		\caption{\textbf{Working principle of the approach used to generate single-phonon states and verify their nonclassicality.} The first step (\textbf{left}) starts with a mechanical oscillator in its quantum ground state, followed by pumping the optomechanical cavity with a blue-detuned pulse. The resonator is excited to a single-phonon state with a probability $p_b=1.2$\% through the optomechanical interaction, which is accompanied by the emission of a photon on resonance with the cavity. The detection of such a photon in a single-photon detector (indicated by the ``Click") allows us to post-select on a purely mechanical Fock state. To verify the quantum state that we created, a red-detuned read pulse is sent onto the optomechanical cavity in the second step (\textbf{right}), which performs a partial state transfer between the optics and the mechanics. With a probability of $p_r=32.5$\%, the mechanical system's excitation is converted into a photon on cavity resonance, returning the mechanics to its ground state. The photon is sent onto a beamsplitter, where we measure the second-order intensity correlation function $g^{(2)}$ by using a pair of single-photon detectors. $g^{(2)}(0)<1$ confirms the nonclassicality of the generated phonon states. The insets show the equivalent energy level diagrams of the processes.}
		\label{Fig:1}
	\end{center} 
\end{figure*}

Over the past decade, motional degrees of freedom (phonons) of solid-state devices have emerged as a quantum resource. Quantum control of phonons was pioneered in the field of trapped ions~\cite{Blatt2008}, where single excitations of the motion of the ions are manipulated through laser light. These single-phonon states have been used for fundamental studies of decoherence~\cite{Leibfried1996} and for elementary transduction channels in quantum gates for universal quantum computing~\cite{Cirac1995}. Cavity optomechanics~\cite{Aspelmeyer2014} has successfully extended these ideas to optically controlling the collective motion of solid-state mechanical systems. It has allowed for remarkable progress in controlling solid-state phonons at the quantum level, including sideband cooling into the quantum ground state of motion~\cite{Teufel2011b,Chan2011}, the generation of quantum correlated states between radiation fields and mechanical motion~\cite{Palomaki2013,Lee2011,Riedinger2016}, and the generation of squeezed motional states~\cite{Wollman2015,Pirkkalainen2015,Lecocq2015}.

So far single-phonon manipulation of micromechanical systems has exclusively been achieved through coupling to superconducting qubits~\cite{OConnell2010,Chu2017,Reed2017}, and optical control has been limited to the generation of quantum states of bipartite systems~\cite{Lee2011,Lee2012,Riedinger2016}. Here we demonstrate all-optical quantum control of a purely mechanical system, creating phonons at the single quantum level and unambiguously showing their nonclassical nature. We combined optomechanical control of motion and single-phonon counting techniques~\cite{Cohen2015,Riedinger2016} to probabilistically generate a single-phonon Fock state from a nanomechanical device. Implementing Hanbury Brown and Twiss interferometry for phonons~\cite{Cohen2015,Riedinger2016} (Figure~\ref{Fig:1}) allowed us to probe the quantum mechanical character of single-phonons without reconstructing their states. We observed $g^{(2)}(0)<1$, which is a direct verification of the nonclassicality of the optomechanically generated phonons, highlighting their particle-like behavior.

\begin{figure}[t]
	\begin{center}
		\includegraphics[width=1.0\columnwidth]{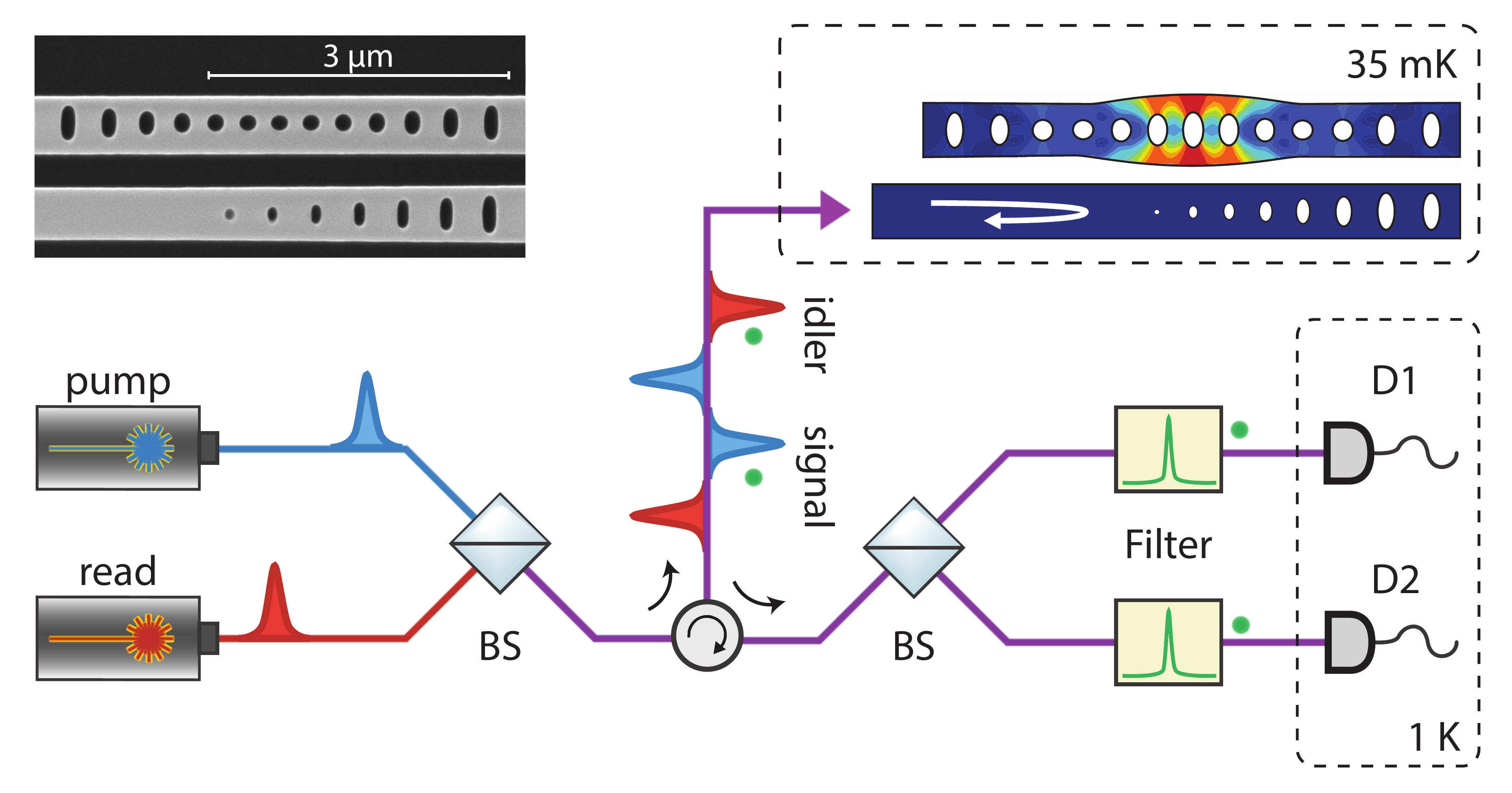}
		\caption{\textbf{Sketch of the experimental setup used to measure the intensity autocorrelation function $g^{(2)}$ of phonons.} Blue-detuned pump pulses are sent into the optomechanical cavity, which is kept at 35~mK. With a small probability $p_b$, the optomechanical interaction creates a single excitation of the mechanical mode at 5.25~GHz (idler) and at the same time emits a signal photon on resonance with the cavity. The original optical pump field is then filtered and only the signal photon created in the optomechanical down-conversion process is detected in one of the single-photon detectors ($D1$ or $D2$). With a time delay $t_d$, a red-detuned read pulse is sent into the device, converting any mechanical idler excitation into an idler photon, which again is filtered from the original pump. Conditioned on the detection of a signal photon, we measure the $g^{(2)}$ of the idler photons. Because the red-detuned pulse is equivalent to a state-swap interaction, the $g^{(2)}$ function that we obtain for the photons is a direct measure of the $g^{(2)}$ function of the phonons in the mechanical oscillator. The inset in the top left corner shows a scanning electron microscope image of the device (top) next to a waveguide (bottom). BS, beamsplitter.}
		\label{Fig:2}
	\end{center} 
\end{figure}

Our optomechanical crystal~\cite{Chan2011} consists of a microfabricated silicon nanobeam patterned so that it simultaneously acts as a photonic and phononic resonator (Figure~\ref{Fig:2}). The resulting optical and mechanical modes couple through radiation pressure and the photoelastic effect so that a displacement equivalent to the zero-point fluctuation of the mechanical mode leads to a frequency shift of the optical mode by $g_0/2\pi=869$~kHz ($g_0$:\  optomechanical coupling rate). The optical resonance has a wavelength $\lambda=1554.35$~nm and a critically coupled total quality factor $Q_o=2.28\times10^5$ (cavity energy decay rate $\kappa/2\pi=846$~MHz), whereas the mechanical resonance has a frequency of $\omega_m/2\pi=5.25$~GHz and a quality factor of $Q_m=3.8\times10^5$. The device is placed in a dilution refrigerator with a base temperature of $T=35$~mK. When the device is thermalized, its high frequency guarantees that the mechanical mode is initialized deep in its quantum ground state~\cite{Riedinger2016}.

We utilized two types of linearized optomechanical interactions -- the parametric down-conversion and the state swap -- which can be realized by driving the system with detuned laser beams in the limit of weak coupling ($g_0\sqrt{n_c}\ll \kappa$, where $n_c$ is the intracavity photon number) and resolved sidebands ($\kappa\ll \omega_m$)~\cite{Aspelmeyer2014}. The parametric down-conversion interaction has the form $H_{dc}=\hbar g_0 \sqrt{n_c} (\hat{a}^\dag \hat{b}^\dagger+\hat{a}\hat{b})$ where $\hbar$ is the reduced Planck constant; $\hat{b}^\dag$  and $\hat{b}$ are the phononic creation and annihilation operators, respectively; and $\hat{a}^\dag$ and $\hat{a}$ are the respective photonic operators. This interaction is selectively turned on by detuning the laser frequency $\omega_L$ to the blue side of the cavity resonance $\omega_c$ ($\omega_L=\omega_c+\omega_m$). $H_{dc}$ drives the joint optical and mechanical state, initially in the ground state, into the state $|\psi\rangle_{om}\propto|00\rangle+p_b^{1/2}|11\rangle+p_b|22\rangle+\mathcal{O}(p_b^{3/2})$. For low excitation probabilities $p_b\ll 1$, higher-order terms can be neglected so that the system can be approximated as emitting a pair consisting of a resonant signal photon and an idler phonon with a probability $p_b$~\cite{Hofer2011}. Detection of the signal photon emanating from the device heralds a single excitation of the mechanical oscillator $|\psi\rangle_m\approx|1\rangle$, in close analogy to heralded single-photons from spontaneous parametric down-conversion. To read out the phonon state, we send in another laser pulse that is now red-detuned from the cavity resonance by $\omega_m$ ($\omega_L=\omega_c-\omega_m$). This realizes a state-swap interaction $H_{swap}=\hbar g_0 \sqrt{n_c} (\hat{a}^\dag \hat{b}+\hat{a}\hat{b}^\dag)$, which transfers the mechanical state to the optical mode with efficiency $p_r$. We can therefore use the scattered light field from this ``read" operation to directly measure the second-order intensity correlation function $g^{(2)}$ of the mechanical oscillator mode, which is defined as
\begin{equation}
g^{(2)}(\tau)=\langle \hat{b}^\dag(0)\hat{b}^\dag(\tau) \hat{b}(\tau)\hat{b}(0)\rangle/\langle\hat{b}^\dag(0)\hat{b}(0)\rangle \langle\hat{b}^\dag(\tau)\hat{b}(\tau)\rangle,
\end{equation}
where $\tau$ is the time between the first and the second detection event. Like for any other bosonic system, $g^{(2)}(0)>1$ means that the phonons exhibit super-Poissonian (classical) behavior, whereas $g^{(2)}(0)<1$  is direct evidence of the quantum mechanical nature of the state and implies sub-Poissonian phonon statistics~\cite{Davidovich1996}.

We implemented the experimental approach (Fig.~\ref{Fig:1}) by repeatedly sending a pair of optical pulses, the first one blue-detuned [pump pulse, full width at half maximum (FWHM) $\approx$ 32~ns] and the second one red-detuned (read pulse, FWHM $\approx$ 32~ns) with a fixed repetition period $T_r = 50$~$\mu$s. Photons generated through the optomechanical interactions were reflected back from the device and analyzed by a Hanbury Brown and Twiss interferometer using two superconducting nanowire single-photon detectors (SNSPDs). We set the mean pump pulse energy to 27~fJ so that $p_b = 1.2\%$~\cite{SM}. Detection of resonant (signal) photons created by this pulse heralds the preparation of the mechanical oscillator in a single-phonon Fock state, in principle with a probability of 98.8\%. Owing to a small amount of initial thermal phonons and residual absorption heating, a fraction of unwanted phonons were incoherently added to the quantum states that we prepared~\cite{Riedinger2016}. After each pump pulse, a red-detuned read pulse was sent to the device with a programmable delay $t_d$, reading out phonons stored in the device by converting them into photons on resonance with the cavity. The mean read pulse energy is set to 924~fJ, corresponding to a state-swap efficiency $p_r\approx32.5$\%. Taking into account subsequent optical scattering losses, this yields an absolute quantum efficiency for the detection of phonons of 0.9\%~\cite{SM}. Last, the pulse repetition period of $T_r=50~\mu$s, which is long compared with the mechanical damping time of 11~$\mu$s, provides ample time for dissipating any excitation or unwanted heating generated by optical absorption. This ensured that each experimental cycle started with the mechanical mode well in the quantum ground state. The pulse sequence was repeated more than $7\times10^9$ times to acquire enough statistics. Conditioned on heralding events from detector $D1$ by the blue-detuned pulses, we analyzed the coincidence detection probability of photons at $D1$ and $D2$ that are transferred from phonons by the swap operation.

\begin{figure*}[t]
	\begin{center}
		\includegraphics[width=1.9\columnwidth]{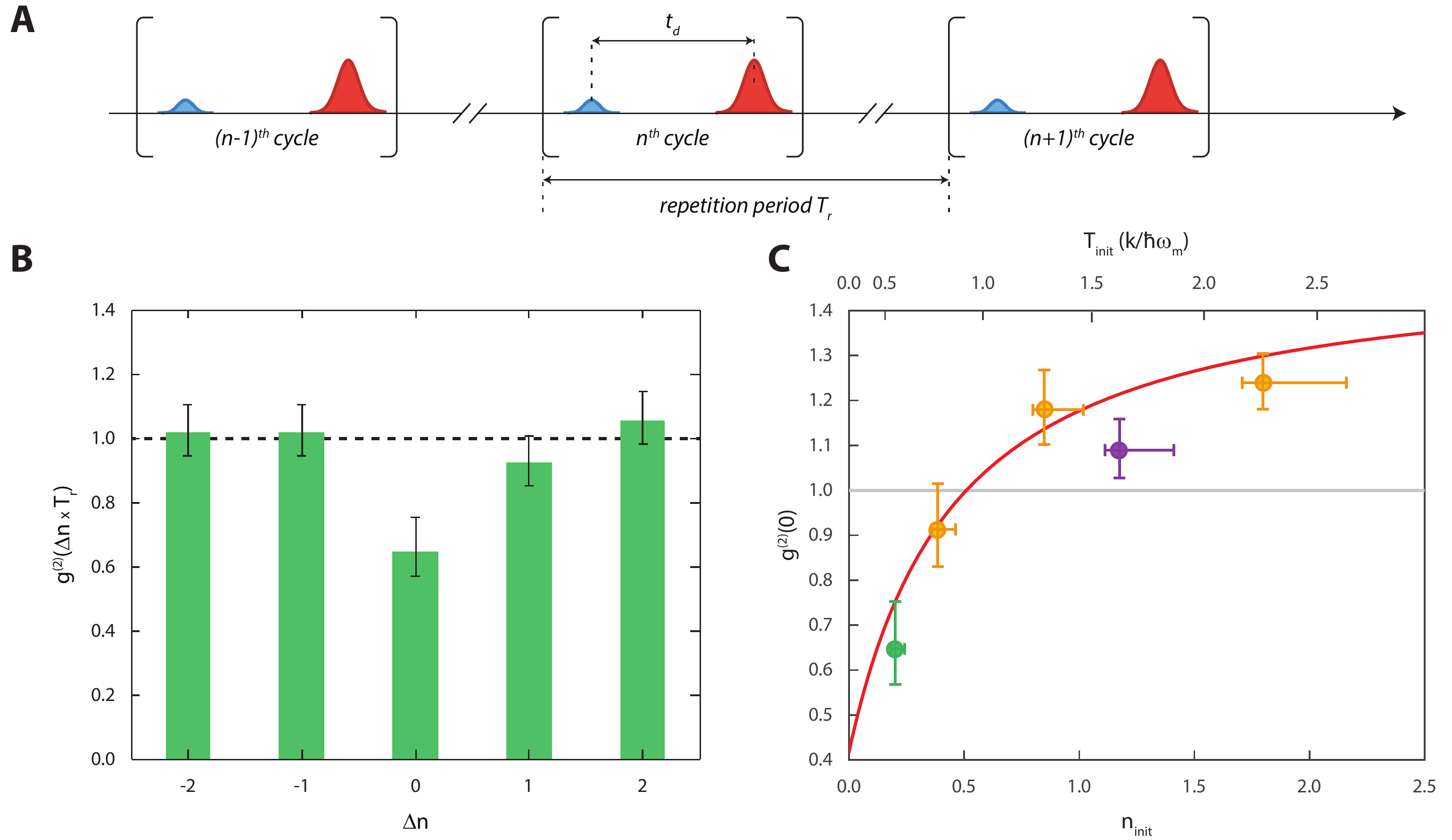}
		\caption{\textbf{Experimental single-phonon creation and HBT interferometry. A} Pulse sequence used in the experiments. Each cycle consists of a blue-detuned pump pulse and a subsequent red-detuned read pulse delayed by $t_d$. The pulse sequence is repeated with the period $T_r$. Both $t_d$ and $T_r$ can be adjusted. \textbf{B} The measurement result of the second-order correlation function $g^{(2)}(\tau=\Delta n\times T_r)$ of the heralded phonons, with $g^{(2)}(0)<1$ being a direct measure of their nonclassicality. In this measurement, we set $t_d=115$~ns and $T_r=50~\mu$s. $g^{(2)}(\Delta n\times T_r)$ with $\Delta n\neq0$ depicts the correlations between phonons read from separate pulse sequences with the cycle difference of $\Delta n$. Whereas phonons from independent pulses show no correlation [$g^{(2)}(\Delta n\times T_r;\Delta n\neq0)\approx 1$], those from the same read pulse are strongly anticorrelated [$g^{(2)}(\tau=0)=0.65^{+0.11}_{-0.08}$]. \textbf{C} The influence of an incoherent phonon background on the $g^{(2)}(0)$ of the generated mechanical states. Several measurements are plotted for a range of different effective initial temperatures of the nanomechanical oscillator. The first data point (green) was taken with a delay $t_d=115$~ns and a repetition period $T_r=50~\mu$s. We control the initial mode occupation $n_{init}$ (initial mode temperature $T_{init}$) by using the long lifetime of the thermally excited phonons stemming from the delayed absorption heating by pump and read pulses. This allows us to increase $n_{init}$ while keeping the bulk temperature and properties of the device constant, causing an increase in $g^{(2)}(0)$, as the state becomes more thermal. The red line shows the simulated $g^{(2)}(0)$ as discussed in~\cite{SM}. For technical reasons all data points (yellow and purple) except the leftmost (green) were taken with $t_d=95$~ns. In addition, the second from the right (purple) was taken at an elevated bath temperature of $T_{bath}=160$~mK.}
		\label{Fig:3}
	\end{center} 
\end{figure*}

In our first experiment, we set $t_d=115$~ns and measured $g^{(2)}(0)$ of the heralded phonons. One of our SNSPDs, $D2$, exhibited a longer dead time than $t_d$~\cite{SM} and we therefore only used photon counts from $D1$ for heralding the phonon states. From these measurements, we obtained a $g^{(2)}(0)$ of $0.65^{+0.11}_{-0.08}$ (Figure~\ref{Fig:3}C), demonstrating a nonclassical character of the mechanical state.

The observed $g^{(2)}(0)$ of 0.65 is considerably higher than what we expect in the ideal case $g^{(2)}_{ideal}(0) \approx 4\times p_b = 0.045$~\cite{SM}. We attribute this to heating induced by the absorption of the pump and read pulses. Although a detailed physical mechanism for the absorption and subsequent heat transfer into the mechanical mode is still a subject of study~\cite{Riedinger2016}, the influx of thermal phonons $\dot n_{abs}$ caused by the absorption of drive laser pulses can be experimentally deduced from the (unconditional) photon count rates generated by the read pulses~\cite{SM}. Including an estimation of the initial thermal phonon number $n_{init}$, which is likewise inferred from the unconditional photon counts associated with the pump and read pulses, we constructed a theoretical model that predicts $g^{(2)}(0)$ as a function of $p_b$, $n_{init}$, and $\dot n_{abs}$. Given the measured $n_{init}\approx0.20$ and $\dot n_{abs}$~\cite{SM} within the read pulse, our model predicts $g^{(2)}(0)\approx 0.76$, which is consistent with the experimental value.

To further probe the effect of thermal phonons, we performed a set of experiments with reduced repetition periods $T_r$, while keeping the other settings for the pump pulses the same. This effectively increases $n_{init}$, because the absorbed heat does not have enough time to dissipate before the next pair of pulses arrives. As expected, as $T_r$ was reduced, we observed an increase in $g^{(2)}(0)$. With the measured $n_{init}$ and $\dot n_{abs}$ from the same data set, we can plot the predicted $g^{(2)}(0)$ values. The experimental values and theoretical bounds on $g^{(2)}(0)$ are in good agreement (Fig.~\ref{Fig:3}).

We also measured $g^{(2)}(0)$ for $t_d=350$~ns and found that it increased to $0.84^{+0.07}_{-0.06}$. This increase is consistent with previously observed delayed heating effects of the absorption~\cite{Riedinger2016} and is in good agreement with the theoretical prediction of 0.84. Even for these longer delays, the value is still below 1, demonstrating the potential of our device as a single-phonon quantum memory on the time scale of several hundred nanoseconds.

We experimentally demonstrated the quantum nature of heralded single-phonons in a nanomechanical oscillator by measuring their intensity correlation function $g^{(2)}(0)<1$. The deviation from a perfect single-phonon state can be modeled by a finite initial thermal occupation and additional heating from our optical cavity fields. We achieved conversion efficiencies between phonons and telecom photons of more than 30\%, only limited by our available laser power and residual absorption. Full state reconstruction of the single-phonon state, as demonstrated with phononic states of trapped ions~\cite{Leibfried1996}, should be realizable with slightly improved read-out efficiency and through homodyne tomography. The demonstrated fully optical quantum control of a nanomechanical mode, preparing sub-Poissonian phonons, shows that optomechanical cavities are a useful resource for future integrated quantum phononic devices, as both single-phonon sources and detectors. They are also an ideal candidate for storage of quantum information in mechanical excitations and constitute a fundamental building block for quantum information processing involving phonons. Some of the potential applications include quantum noise--limited, coherent microwave-to-optics conversion, as well as studying the quantum behavior of individual phonons of a massive system. 

\vspace{-0.5cm}
\section*{Acknowledgments}
We thank V. Anant, J. Hofer, C. Loeschnauer, R. Patel, and A. Safavi-Naeini for experimental support and K. Hammerer for helpful discussions. We acknowledge assistance from the Kavli Nanolab Delft, in particular from M. Zuiddam and C. de Boer. This project was supported by the European Commission under the Marie Curie Horizon 2020 initial training programme OMT (grant 722923), Foundation for Fundamental Research on Matter (FOM) Projectruimte grants (15PR3210, 16PR1054), the Vienna Science and Technology Fund WWTF (ICT12-049), the European Research Council (ERC CoG QLev4G, ERC StG Strong-Q), the Austrian Science Fund (FWF) under projects F40 (SFB FOQUS) and P28172, and by the Netherlands Organisation for Scientific Research (NWO/OCW), as part of the Frontiers of Nanoscience program, as well as through a Vidi grant (016.159.369). R.R. is supported by the FWF under project W1210 (CoQuS) and is a recipient of a DOC fellowship of the Austrian Academy of Sciences at the University of Vienna.

%\bibliography{../../Mirror}

%\clearpage

\setcounter{figure}{0}
\renewcommand{\thefigure}{S\arabic{figure}}
\setcounter{equation}{0}
\renewcommand{\theequation}{S\arabic{equation}}

\section*{Supplementary Materials}

\subsection*{Optomechanical devices}
The optomechanical device is fabricated from a silicon on insulator wafer (Soitec) with a device layer of 250~nm thickness on top of a 3~$\mu$m buried oxide layer. We pattern our chips with an electron beam writer and transfer the structures into the silicon layer in a reactive ion etcher using a $\mathrm{SF_6/O_2}$ plasma. One of the sides of the chip is removed to allow for in-plane access to the lensed fiber couplers. After the resist is removed, the device layer is undercut in 40\% hydrofluoric acid. An additional cleaning step using the so-called RCA method~\cite{Kern1970} is performed to remove organic and metallic residuals. The final step is a dip in 2\% hydrofluoric acid to remove the oxide layer formed by the RCA cleaning and to terminate the silicon surface with hydrogen atoms.

Unlike in previous device designs~\cite{ChanPhD}, we do not use an additional phononic shield around the optomechanical structure as this unnecessarily increases the re-thermalization time and therefore reduces the achievable repetition rate of our experiment~\cite{Riedinger2016}. We reduce the mechanical quality factors of the designed structures further by offsetting the photonic crystal holes laterally from the center of the beam by 30~nm~\cite{Patel2017}. This yields a measured quality factor of $Q_m=3.8\times10^5$ at mK temperatures, while otherwise such structures exhibit Q's beyond $10^7$.

In order to find particularly good devices on a chip, we characterize them  in a pump-probe experiment at cryogenic temperatures (35mK) and select devices with optimal mechanical Q and low optical absorption. We then perform cross-correlation measurements of the photon-phonon pairs scattered by the pump pulse, while varying the repetition period $T_r$, pump excitation probability $p_b$ and state-swap efficiency $p_r$. This short two-fold coincidence measurement ($\sim$1h) allows us to predict the expectation value of the three-fold coincidence autocorrelation measurement~\cite{Chou2004,Galland2014,Riedinger2016} as well as the time required to obtain enough statistics for a targeted confidence interval. We chose a parameter set, which allows for a statistically significant (p-value $< 0.001$, see below) demonstration of intensity anticorrelations ($g^{(2)}(0)<1$) of the phononic state within a realistic measurement time ($\sim$100h).

\begin{figure}[h!]
	\centering
	\includegraphics[width=.9\columnwidth]{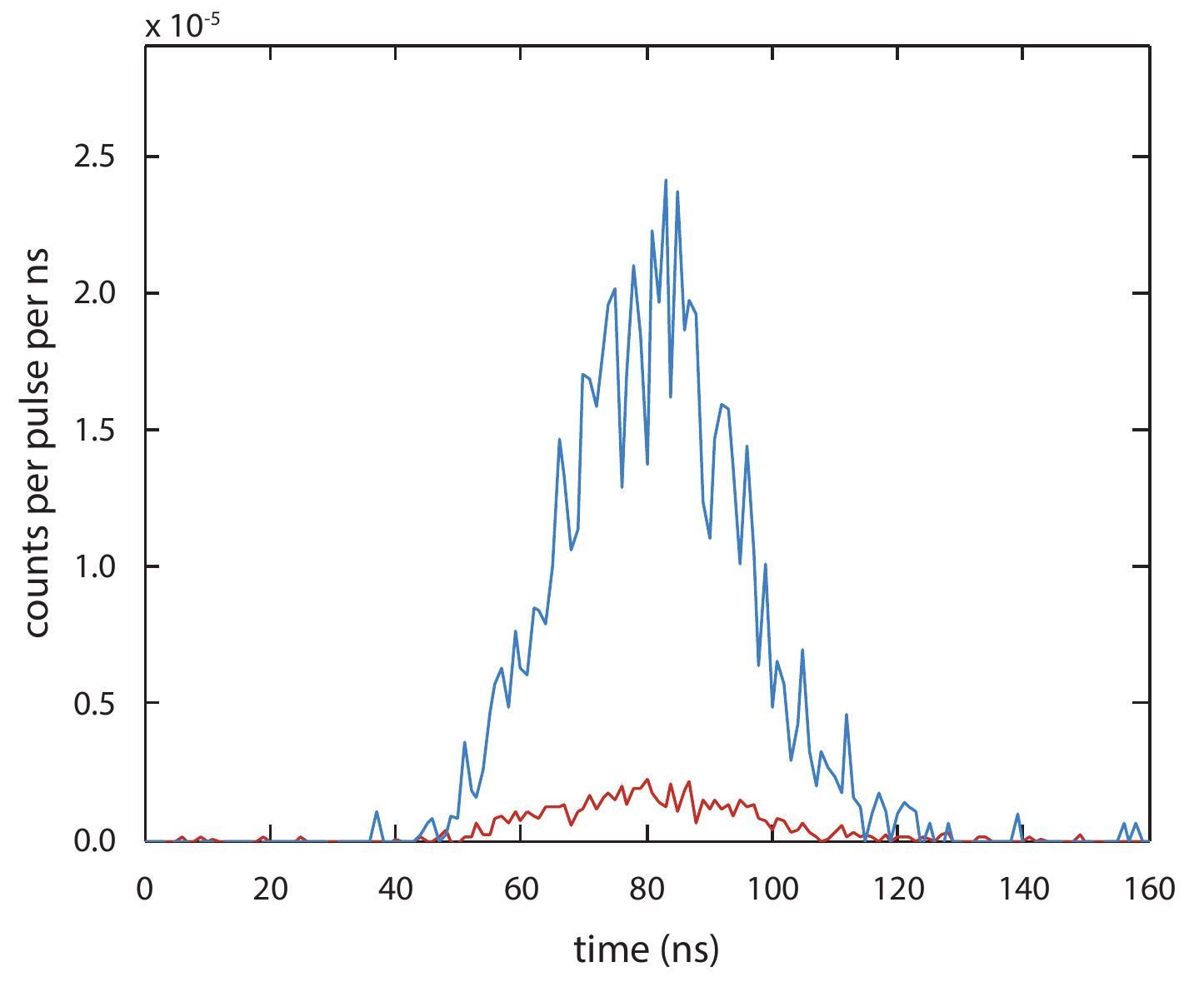}
	\caption{A sideband asymmetry measurement is performed to extract the optomechanical coupling rate $g_{0}$. For a detailed explanation of the measurement see the text below. We plot the detected photon counts per pulse repetition per nanosecond for a blue-detuned (blue) and a red-detuned beam (red). Unwanted contributions from leaked pump photons and detector dark counts are independently measured and subtracted. Integrating over the whole detection window (i.e.\ from 30 to 150~ns in the plot) gives the photon counting probabilities $C_{b}$ and $C_{r}$, respectively. From this data, we extract the optomechanical coupling rate $g_{0}/2\pi = 869$~kHz.}
	\label{SM:Fig:1}
\end{figure}

\subsection*{Detection efficiency}
We calibrate the total detection efficiencies of optomechanically generated cavity photons ($\eta_{i}; i=1,2$) by performing a series of independent measurements. First, the fiber-to-device coupling efficiency ($\eta_{fc}=0.48$) is measured by sending light with known power to the photonic crystal and then measuring the reflected power. The extraction efficiency of cavity photons $\eta_{dev}$ is obtained from the device impedance ratio, $\eta_{dev}=\kappa_{e}/\kappa$, where $\kappa_e$ is the external cavity energy decay rate. These values are extracted from the visibility and the linewidth of the optical resonance scan, and we find $\eta_{dev}=0.5$. Furthermore, we measure the efficiency of detecting photons coming from the device for each SNSPD. We launch weak, off-resonant optical pulses with an average of 5.14 photons to the device and measure the photon count rate of each SNSPD. This measurement gives the quantities $\eta_{fc}^{2}\times\eta_{trans,i}\times\eta_{QE,i}$, where $\eta_{trans,i}$ is the transmission efficiency of the detection path to each SNSPD, while $\eta_{QE,i}$ is their quantum efficiency. As $\eta_{fc}$ is measured independently, $\eta_{trans,i}\times\eta_{QE,i}$ can be calculated from these results. Finally, this allows us to obtain $\eta_{i}=\eta_{dev}\times\eta_{fc}\times\eta_{trans,i}\times\eta_{QE,i}$, which are $\eta_{1}=1.16$\% and $\eta_{2}=1.50$\%, respectively.

\subsection*{Optomechanical coupling rate}
In order to calibrate the optomechanical coupling rate $g_0$ between the cavity field and the mechanical mode, we perform a measurement similar to sideband thermometry~\cite{Meenehan2015,Riedinger2016} (see Figure~\ref{SM:Fig:1}). In this measurement, pairs of pump and probe pulses are sent to the device with a repetition period of $T_r=100~\mu$s. For this pulse sequence, a blue-detuned pump pulse (190.62~fJ) is sent to intentionally heat the device's mechanical mode, followed by the probe pulse (55.16~fJ) with a long delay of 99.825~$\mu$s. We perform two sets of such repetitive measurements, one with red-detuned and the other with blue-detuned probe pulses. From each measurement, we acquire a photon counting probability for the probe pulses, $C_{r}$ (red-detuned) and $C_{b}$ (blue-detuned). These can be expressed as $C_{r}=(\eta_{1}+\eta_{2})\times p_{r}\times n_{th}$ and $C_{b}=(\eta_{1}+\eta_{2})\times p_{b}\times (1+n_{th})$ in the limit of $p_b\ll 1$ and $p_r\ll 1$, where $p_b$ and $p_r$ are equivalent to the photon-phonon pair excitation probability and state-swap efficiency as introduced in the main text. $p_b$ and $p_r$ can be explicitly written as
\begin{equation}
p_{b} = \exp(\kappa_e/\kappa \left[4g_0^2 E_{p}/\hbar\omega_c(\omega_m^2+(\kappa/2)^2)\right])-1,
\end{equation}
\begin{equation}
p_{r} = 1-\exp(-\kappa_e/\kappa \left[4g_0^2 E_{p}/\hbar\omega_c(\omega_m^2+(\kappa/2)^2)\right]),
\end{equation}
where $E_p$ is the total energy of the incident laser pulses and all the other terms as defined in the main text. We find that $C_{r}=0.0064$\% and $C_{b}=0.0697$\%. From these values we extract $n_{th}=0.104$, $p_{r}=2.32\%\ll 1$ and $p_{b}=2.37\%\ll 1$, which allows us to directly obtain $g_{0}/2\pi = 869$~kHz, in good agreement with our simulated value~\cite{Riedinger2016}. With the calibrated value of $g_0$, the scattering probabilites $p_r$ and $p_b$ can now be directly set by simply choosing the appropriate pulse energies. The average phonon occupation of the mechanical oscillator $n_{th}$ can also be obtained by measuring the count rates with predetermined values of $p_b$ and $p_r$, without requiring sideband thermometry.

\subsection*{Data analysis}
The second order autocorrelation function is defined as
\begin{equation}
g^{(2)}(t_1,t_2)=\frac{\left\langle : \hat{N}(t_1)\hat{N}(t_2):\right \rangle}{\left\langle\hat{N}(t_2)\right \rangle \left\langle\hat{N}(t_1)\right \rangle},
\end{equation}
where $\hat{N}(t)=\hat b ^\dagger (t) \hat b (t)$ is the phonon number operator of the mechanical mode at time $t$ after the start of the pulse sequence, and $:\; :$ is the notation for time and normal ordering of the operators. The mechanical mode is measured by the optical read pulses and the signal, i.e.\ the scattered photons, are filtered before they are detected by SNSPDs. Consequently, the observed detection events are averaged by the optical filters and weighted with the envelope of the read pulse $n_p(t)$, which holds for the weak coupling (i.e.\ adiabatic) regime. Further, to gain enough statistics, the events associated with the read pulse are integrated. We define the time interval $\left[t_a, t_b\right]$, containing the effective pulse shape $p(t)$, which is obtained from the actual pulse envelope $n_p(t)$ and the filter transfer function. This allows us to express the observed autocorrelation function  
\begin{equation}
g^{(2)}_{obs}(\tau)=\frac{\int_{t_a}^{t_b} dt_1 \int_{t_a+\tau}^{t_b+\tau} dt_2 p(t_1)p(t_2)\left\langle : \hat{N}(t_1)\hat{N}(t_2):\right \rangle}{\left(\int_{t_a}^{t_b} dt_1 p(t_1)\left\langle\hat{N}(t_1)\right\rangle\right) \left(\int_{t_a+\tau}^{t_b+\tau} dt_2 p(t_2)\left\langle\hat{N}(t_2)\right \rangle\right)},
\end{equation}
for a delay $\tau$ between two phonon measurements. This averaging does not influence the validity  of the statements about sub-poissonian statistics and nonclassicality of the mechanical state. Strictly speaking, within the averaging window there is a randomization of the phonon statistics due to damping and heating. Thus, a small regression towards $g^{(2)}=1$ is the expected. Due to the short averaging time, this effect is negligible compared the other uncertainties and systematic effects described below, such that we can safely assume $g^{(2)}(\tau)\equiv g^{(2)}(t_d, t_d+\tau)\approx g^{(2)}_{obs}(\tau)$, with the effective delay of the read pulse $t_d=(t_a+t_b)/2$.

A Hanbury Brown and Twiss setup with two single-photon detectors $D1$ and $D2$ with low count rates allows to measure this second order autocorrelation~\cite{Hofer2011,Galland2014}. Specifically, for $\tau=0$, this expression reduces to the cross-correlation between those detectors
\begin{equation}
g^{(2)}(0)\approx g^{(2)}_{E_{1},E_{2}}=P(E_1\cap E_2)/P(E_1)P(E_2),\label{eq:g2prob}
\end{equation}
where $P(X)$ describes the probability of the occurrence of event $X$, and $E_n$ is a detection event at detector $D_n$ ($n=1,2$) during the time interval $\left[t_a, t_b\right]$. In this notation, it can easily be seen that a rescaling of the detection efficiency of either detector drops out of the expression. Consequently, $g^{(2)}$ is independent of losses in the optical path or the fidelity of the state transfer by the read pulse. However, the value of $g^{(2)}$ can be changed by measurement noise, in our case dominated by false positive detection events (caused by electronic noise, stray light or leaked pump photons). In our setup this gives a negligible systematic error $\delta g^{(2)} = g^{(2)}_{E_{1},E_{2}}-g^{(2)}(0)$ of $0<\delta g^{(2)}<3\times 10^{-4}$.
If the state-swap is seen as part of the measurement, heating of the mechanical state by optical absorption of pump photons within the device can also be interpreted as measurement noise. The effect on $g^{(2)}$ depends strongly on the initial effective temperature of the mechanical mode, so that we cannot give a general number for the systematic error. From simulations, we deduce that it spans from about $0<\delta g^{(2)}_{abs}<0.17$ for the lowest temperature measurement to $0<\delta g^{(2)}_{abs}<0.02$ for the highest initial temperature. The absorption heating in combination with dead time of the SNSPDs, additionally causes a systematic error of $0<\delta g^{(2)}_{dt}<0.03$, which is described in detail in the following section. As the heating related effects can also be considered to be part of the actual mechanical state and the other effects are much smaller than the statistical uncertainties, all $g^{(2)}$ values presented in this work are not corrected for these systematic errors.
With all $\delta g^{(2)}>0$, the presented values are upper bounds to the noise free auto-correlation of the mechanical state.

To estimate the statistical uncertainty of our measurement, we use the likelihood function based on a binomial distribution of photon detection events in the limit of low probabilities. The experimentally measurable values for $g^{(2)}_{E_{1},E_{2}}(0)$ are the maximum likelihood values
\begin{equation}
\bar{g}^{(2)}_{E_{1},E_{2}}\equiv \frac{C(E_{1}\cap E_{2})/N}{(C(E_{1})/N)(C(E_{2})/N)} \approx g^{(2)}_{E_{1},E_{2}},\label{eq:g2likelyhood}
\end{equation}
where $C(E_{1})$ ($C(E_{2})$) is the number of counts registered at detector $D1$ ($D2$) and $C(E_{1}\cap E_{2})$ is the number of co-detection events at both detectors, all conditioned on heralding events (i.e.\ detection events from earlier pump pulses). $N$ refers to the number of such heralding events. In our experiment, the uncertainty of $g^{(2)}_{E_{1},E_{2}}$ is dominated by that of $\bar{P}(E_{1}\cap E_{2}) \equiv C(E_{1}\cap E_{2})/N$ , i.e.\ the estimated probability of $P(E_{1}\cap E_{2})$, as $E_1\cap E_2$ is the rarest event among all the other events. Therefore, we use the likelihood function of $P(E_{1}\cap E_{2})$ to determine the confidence interval of the given values $g^{(2)}(0)=\bar g^{(2)}_{E_{1},E_{2}}\hspace{1pt}^{+\sigma_+}_{-\sigma_-}$, such that the likelihood of the actual value of $g^{(2)}_{E_{1},E_{2}}$ is 34\% to be within $\left[ \bar g^{(2)}_{E_{1},E_{2}}- \sigma_-, \bar g^{(2)}_{E_{1},E_{2}} \right]$ and 34\% to be within $\left[\bar g^{(2)}_{E_{1},E_{2}}, \bar g^{(2)}_{E_{1},E_{2}}+ \sigma_+\right]$. While the low count numbers produce skewed likelihood functions and therefore unequal upper and lower uncertainties $\sigma_\pm$, the counts are high enough such that the rule of thumb of requiring $3\sigma$ for statistical significance (p-value$<0.001$) still holds. Specifically, our null hypothesis is no correlation between the phonons in the oscillator, i.e.\ an actual $g^{(2)}_\textrm{actual}= 1$. For the delay of the read pulse of $t_d=115$~ns and $T=35$~mK, we measured an autocorrelation of $g^{(2)}(0)=0.647^{+0.105}_{-0.079}$. The p-value, i.e.\ the probability of observing this or a more extreme result, given that the null hypothesis of no correlation was true, is $p=P\big(\bar g^{(2)}_{E_{1},E_{2}}\leq 0.647\big|g^{(2)}_\textrm{actual}= 1, N=1.2\times 10^6\big)<7\times 10^{-4}$. In our case this coincides with the probability of falsely rejecting the classical bound $P\big(g^{(2)}_\textrm{actual}(0)\geq 1 \big| \bar g^{(2)}(0)=0.647^{+0.105}_{-0.079}\big)<7\times 10^{-4}$. For the delay of $t_d=370$~ns and $T=35$~mK, we find the p-value $p<0.01$ for the observed $g^{(2)}(0)=0.832^{+0.068}_{-0.058}$, also coinciding with $P\big(g^{(2)}_\textrm{actual}(0)\geq 1 \big| g^{(2)}(0)=0.832^{+0.068}_{-0.058}\big)<0.01$.

\begin{figure}[h!]
	\centering
	\includegraphics[width=.9\columnwidth]{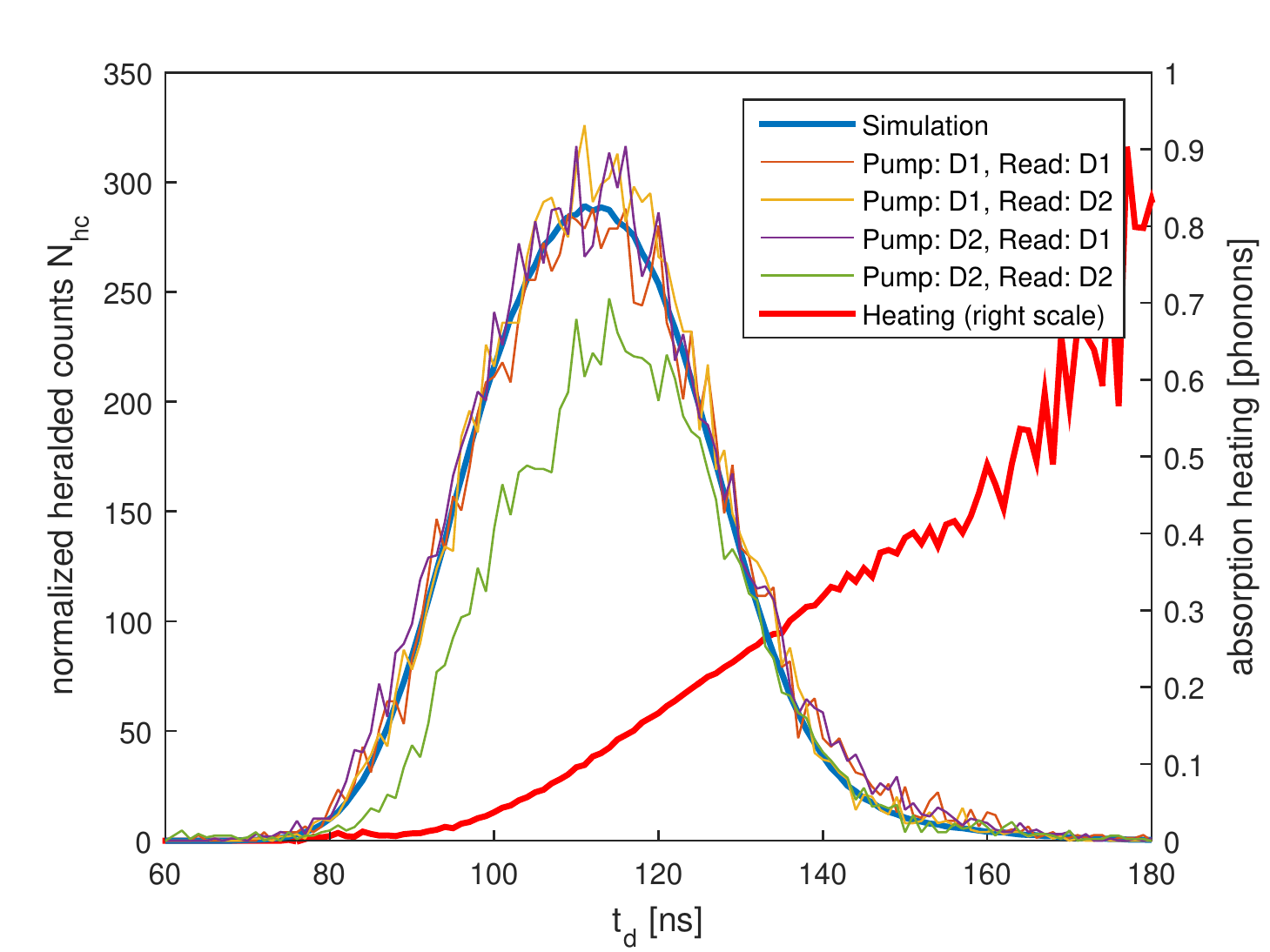}
	\caption{Plotted are the heralded counts for various combinations of detection events:\ 
	In brown, yellow, purple and green are the counts of photons scattered by the read pulse (left axis), heralded on the detection of a photon scattered by the pump pulse. They are normalized to the detection efficiency of events heralded by detector $D1$ and detected by $D2$ (yellow). All combinations involving $D1$ either for heralding or detection match well with the simulated counting distribution (blue). Notably, the combination of heralding and detecting with $D2$ deviates from that. This reduced detection efficiency is caused by the longer dead time of $D2$ compared to $D1$. In addition, the time window over which the pulse averages, is shifted to later times and is more heavily influenced by the cumulative heating $n_{abs}(t)$ (red) by absorbed driving photons from the read pulse. The latter is extracted from the detection of a thermal state and compensated for the optomechanical cooling to obtain the true cumulative number of added phonons $n_{abs}(t)$ (right axis).}
	\label{SM:Fig:2}
\end{figure}

\subsection*{Dead time effects}
In order to reduce the effects of absorption heating on the mechanical state~\cite{Meenehan2014,Meenehan2015,Riedinger2016}, it is important to measure the state as quickly as possible upon its generation by the pump pulse. For the data shown in Figure~\ref{Fig:3}B in the main text, the time delay between the read and the pump pulses is $t_d=115$~ns. For the measurements in Figure~\ref{Fig:3}C, $t_d$ is set to 95~ns, except for the first data point, which represents the result of Figure~\ref{Fig:3}B.

In our measurement scheme, we use the same pair of SNSPDs to herald the generation of nonclassical mechanical states as we use to measure $g^{(2)}(0)$ through the read pulses. After a detection event, the superconducting nanowire is in the normal conducting state and is therefore blind to additional photons arriving during this time, before it cools down and returns into the superconducting state. This so-called dead time for our detectors is nominally on the order of 50 -- 70~ns. If the state is heralded by one of the SNSPDs and its dead time overlaps with the arrival time of the photons from the read pulse, its detection efficiency will be lower than the nominal value. In our experiment this is in fact the case for detector D2 in the measurement of $g^{(2)}(0)$ with $t_d=115$~ns (cf.\ Figure~\ref{SM:Fig:2}).

To first approximation, this should have no influence on the value of $g^{(2)}(0)$ itself, as the detection probability of the heralded pulses enters equation ~(\ref{eq:g2prob}) in the numerator $P(E_1 \cap E_2)$ and the denominator $P(E_2)$ in the same way. However, a detected readout pulse with a seemingly distorted shape as in Figure~\ref{SM:Fig:2} effectively measures the mechanical state slightly later than the nominal delay time $t_d$. Due to absorption heating during the read pulse (cf.\ Figure~\ref{SM:Fig:2}), the mechanical state at later times is corrupted by the influx of thermal phonons $\dot n_{abs}(t)$. Therefore, it will increase the observed value of $g^{(2)}(0)$ towards 2, the value of a thermal state. For this reason we discard heralding events from detector D2 for short delays.

A more detailed analysis allows us to quantify the systematic error by the shorter dead time of detector D1. The minor reduction of the detection efficiency for a $t_d$ of 115~ns leads to $0<\delta g^{(2)}_{dt}<0.01$, which corresponds to an overestimation of our observed value of $g^{(2)}(0)$ by much less than our statistical uncertainty. For $t_d=95$~ns, this effect from detector D1 becomes stronger, resulting in 10\% reduced detection efficiency. However, thanks to an increased thermal background, it only produces a systematic error of around $0<\delta g^{(2)}_{dt}<0.03$, which is again smaller than the statistical uncertainty. As they are negligible in magnitude, we did not account for these systematic errors in the reported values of $g^{(2)}$ in the main text.

\subsection*{Simulation of the correlation function}
To calculate the expected value of $g^{(2)}$ we use the formalism developed by Barchielli~\cite{barchielli_quantum_1987,barchielli_direct_1990}. In order to do this, we require a model of the open-system dynamics of our optomechanical system, describing the coupling to the environment. In addition to the typical assumption that the mechanical system couples to a heat bath of a fixed temperature, we observe in our experiment an additional, time-dependent heating effect that is activated by the strong read pulse. In the absence of a microscopic description of this effect, we adopt a simple phenomenological description and model it as a standard Lindblad dynamics with two parameters $\gamma$ and $n_{bath}$, which can be estimated from the singlefold detection events of the exerimental $g^{(2)}$ data for each $n_{init}$. Note that $n_{bath}$ is not the occupation number determined by the dilution refrigerator and is assumed to be a function of time. In essence, the incoherent (thermal) phonon influx $\dot n_{abs}=\gamma n_{bath}(t)$ is the derivative of the cummulative absorbtion heating and can be extracted from singlefold detection events, with calibrated scattering rates from sideband asymmetry measurements (see above) and knowledge of the envelope of the read pulse (cf.\ Figure~\ref{SM:Fig:2}). The mechanical decay rate $\gamma$ is assumed to be the measured decay rate $\omega_m/Q_m$. The Lindblad dynamics stay identical when coupling to a number of different baths with different coupling strengths, as long as the phonon influx and the mechanical decay rate stay constant. For reasons of simplicity, we therefore work with a single phenomenological phonon influx $\gamma\times n_{bath}(t)$.

The evolution of the optomechanical quantum state $\rho$ under open-system
dynamics can be described by a Lindblad master equation~\cite{lindblad_generators_1976} of the form~\cite{Wilson-Rae2007}
\begin{widetext}
\begin{equation}
\label{eq:1}
\dot{\rho}=\mathcal{L}\rho=-\frac{i}{\hbar}[H_{swap}(t),\rho]+\kappa\mathcal{D}[a]\rho + \gamma (n_{bath}(t)+1) \mathcal{D}[b] \rho+\gamma n_{bath}(t) \mathcal{D}[b^{\dagger}] \rho,
\end{equation}
\end{widetext}
where the time dependence in $H_{swap}$ accounts for the time-dependent drive by
the light pulses. The Lindblad terms,
\begin{equation}
\label{eq:2}
\mathcal{D}[s]\rho=s\rho s^{\dagger}-\frac{1}{2}\Bigl(s^{\dagger}s\rho+\rho s^{\dagger}s\Bigr),
\end{equation}
describe the coupling of the system to its electromagnetic environment (second
term in eq.~(\ref{eq:1})) and the mechanical heat bath with a mean occupation
number $n_{bath}$ (third and fourth term in eq.~(\ref{eq:1})). Below we will write
the formal solution of eq.~(\ref{eq:1}) as $\rho(t)=\mathcal{T}(t,t_0)\rho(t_0)$.

To describe a photon-counting measurement with a quantum efficiency $\eta$, we
iteratively solve the master equation as~\cite{Gardiner2004}
\begin{widetext}
\begin{equation}
\label{eq:4}
\rho(t) = \mathcal{S}(t,t_0)\rho(t_0)+\sum_{m=1}^{\infty}\int_0^tdt_m\dots \int_0^{t_2}dt_1 \mathcal{S}(t,t_m)\mathcal{J}\mathcal{S}(t,t_{m-1})\dots \mathcal{J}\mathcal{S}(t_1,0)\rho(t_0),
\end{equation}
\end{widetext}
where we defined the operator $\mathcal{J}\rho=\eta\kappa\, a\rho a^{\dagger}$
(which corresponds to the emission of one photon from the cavity), and the
propagator $\mathcal{S}$ that solves the effective evolution
$\dot{\mathcal{S}}=(\mathcal{L}-\mathcal{J})\mathcal{S}$. Equation~(\ref{eq:4})
allows for the following interpretation: Assuming that we register $m$ photons
on the photo-detector, the \emph{conditional} state of the system is, up to a
normalizing factor, given by the $m$-th term in the sum above. In case no
photons are registered, the unnormalized conditional state is given by
$\mathcal{S}(t,t_0)\rho(t_0)$ instead. The heralded state of the mechanical
system after the blue-detuned write pulse (for a click of the detector at
time $t_{click}$) is thus given by
\begin{equation}
\label{eq:7}
\rho_{click}=\frac{\mathrm{Tr}_{cav}[\mathcal{T}(t_{click},t_0)\rho(t_0)-\mathcal{S}(t_{click},t_0)\rho(t_0)]}{\mathrm{Tr}[\mathcal{T}(t_{click},t_0)\rho(t_0)-\mathcal{S}(t_{click},t_0)\rho(t_0)]}.
\end{equation}
Note that this state is conditioned on a measurement of \emph{at least} one
phonon. For our case where two-fold events are rare, this effectively reduces to
the first term in the sum in eq.~\eqref{eq:4}, i.e., $\rho_{click}\propto
\mathrm{Tr}_{cav}[a^{\dagger}a \mathcal{S}(t_{click},t_0)\rho(t_0)]$.

As the evaluation of $g^{(2)}$ is computationally expensive, we first
adiabatically eliminate the cavity mode from eq.~(\ref{eq:1}), which is possible
in the weak-coupling limit $g_0\times\sqrt{n_c}\ll \kappa$. For the case of the red-detuned read pulse we find the equation (neglecting the very weak optical-spring effect)~\cite{Wilson-Rae2007}
\begin{equation}
\label{eq:5}
\dot{\rho}_m= [\gamma (n_{bath}+1)+\Gamma_-(t)] \mathcal{D}[b]\rho_m+(\gamma n_{bath}+\Gamma_+(t)) \mathcal{D}[b^{\dagger}]\rho_m,
\end{equation}
for the reduced state of the mechanical system $\rho_m$, with
$\Gamma_{\pm}(t)=2\kappa_{e}/\kappa\times g_0^2 n_c(t)
\mathrm{Re} (\eta_{\pm})$, $\eta_-=2/\kappa$, $\eta_+=2/(\kappa+4i\omega_m)$. In
this approximation the photo-counting measurement at the cavity resonance
frequency is described by $\mathcal{J}(t)\rho=\eta \Gamma_-(t) b \rho b^{\dagger}$.

To calculate $g^{(2)}$ of the photons emitted from the cavity (after heralding)
we need to evaluate time- and normal-ordered expectation values of the cavity
output field, which is readily achieved in this formalism. We find~\cite{barchielli_direct_1990}
\begin{align}
\label{eq:3}
\langle \hat{I}(t_1) \rangle &= \mathrm{Tr}[\mathcal{J}(t_1)\mathcal{T}(t_1,t_{click})\rho_{click}],\\
\label{eq:8}
\langle :\hat{I}(t_1)\hat{I}(t_2): \rangle &= \mathrm{Tr}[\mathcal{J}(t_2)\mathcal{T}(t_2,t_1)\mathcal{J}(t_1)\mathcal{T}(t_1,t_{click})\rho_{click}].
\end{align}
To evaluate equations~\eqref{eq:3} and~\eqref{eq:8}, we expand the mechanical operators in a number basis up to a maximal phonon number of 50. We assume the mechanical system to initially be in a thermal state with a mean phonon number $n_{init}$. In optomechanical crystals, the mechanical damping rate $\gamma$ tends to be a function of the environmental temperature. As $\gamma\times n_{bath}$ is effectively treated as a single parameter $\dot n_{abs}$ and the time scale of the simulation is short compared to the mechanical decay time $1/\gamma$, potential changes of $\gamma$ with the bath temperature by up to one order of magnitude do not influence the simulations significantly.

\begin{figure}[h!]
	\centering
	\includegraphics[width=.9\columnwidth]{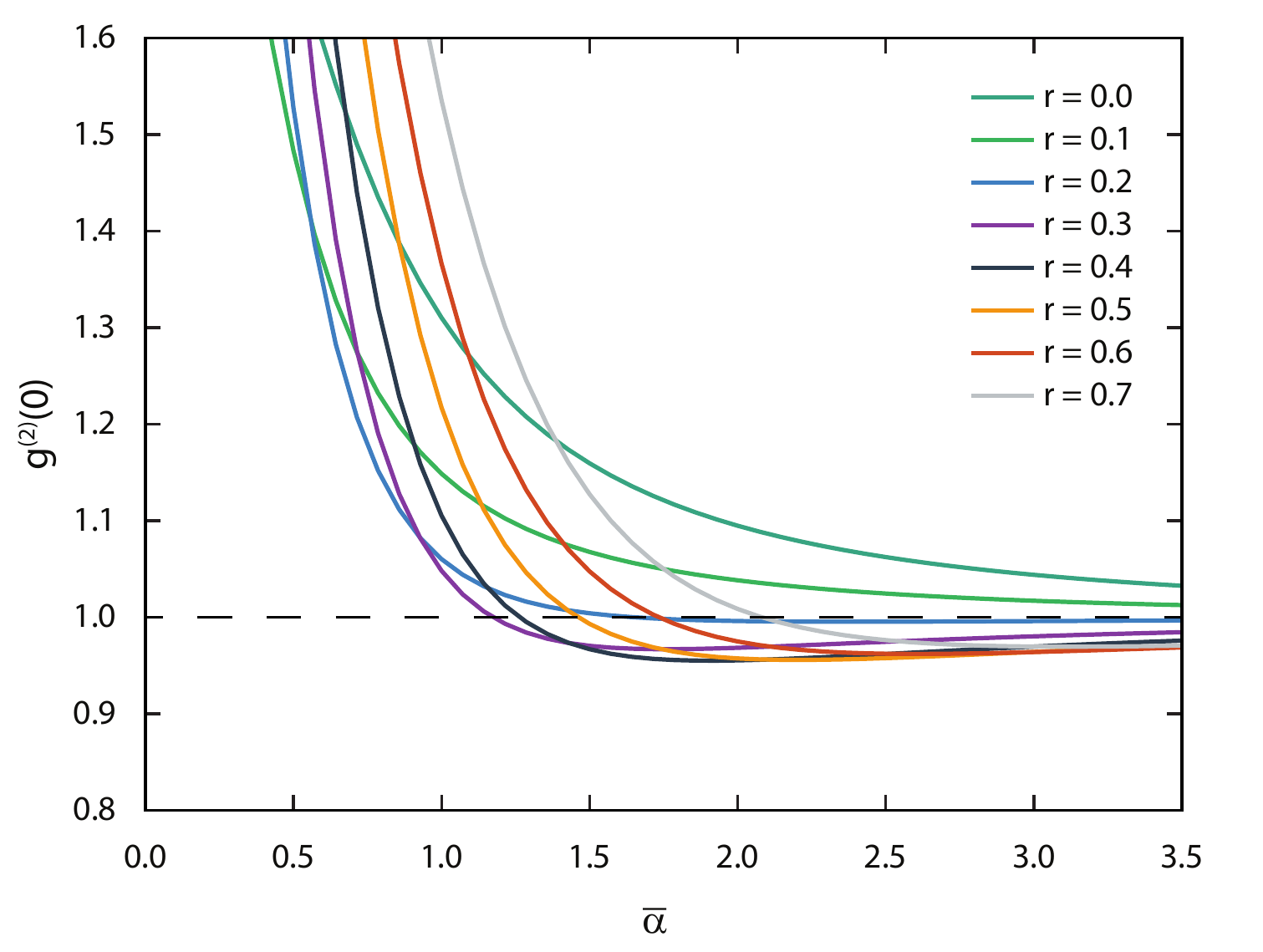}
	\caption{Shown is the numerically calculated $g^{(2)}$ function of a squeezed Gaussian state with an initial thermal occupation of $n_{init}=0.20$ as a function of displacement $\bar{\alpha}$ and squeezing parameter $r$ (color-coded) for $\theta = 2\phi$. Even for the optimal choice of settings $r=0.44$ and $\bar{\alpha}=2.00$, such a model cannot explain our data.}
	\label{SM:Fig:3}
\end{figure}

\subsection*{Nonclassicality and $g^{(2)}$}

The degree of second order coherence $g^{(2)}$ allows to draw various conclusions on the system under investigation. The most prominent use of $g^{(2)}$ is to violate the nonclassicality bound as described in the main text. The physical meaning of this bound can be inferred from the variance of the energy $\hat H=\hbar \omega_m \hat b^\dagger \hat b$ of the free mechanical oscillator.
\begin{equation}
\textrm{Var}(\hat H)=\left\langle \hat H^2 \right\rangle -\left\langle \hat H \right\rangle^2 
=\left( g^{(2)}(0)-1 \right) \left\langle \hat H \right\rangle^2  + \hbar \omega_m \left\langle \hat H \right\rangle.
\label{eq:SM_var}
\end{equation}
For classical physics, i.e.\ not applying canonical quantization, $\langle :\hat H^2:\rangle= \langle\hat H^2\rangle$ and therefore the last term, stemming from the commutation relations, drops out. It immediately follows from $\textrm{Var}(\hat H)\geq 0$ that $g^{(2)}(0)\geq 1$. When using the canonical quantization, we can infer from $g^{(2)}(0)< 1$ that the source had sub-Poissonian phonon statistics. This also classifies the mechanical state as "nonclassical" in the sense that it cannot be represented as an incoherent mixture of coherent states~\cite{Short1983}. The degree of second order coherence $g^{(2)}$ can be used to test against stricter bounds as well~\cite{Filip2013}, some of them depending on the physical system under investigation. For two level systems it is for example important to demonstrate that only a single emitter is present. This can be done by demonstrating $g^{(2)}(0)<0.5$~\cite{Itano1988}. In our case it is sufficient to show $g^{(2)}(0)<1$ as we only have a single macroscopic oscillator by design.

As we can see from equation~\eqref{eq:SM_var}, states possessing a small variance and/or low energies can also exhibit $g^{(2)}(0)<1$. One set of states, which is interesting to exclude is  incoherent mixtures of Gaussian states. In general, linear bosonic systems that involve squeezing can exhibit $g^{(2)}(0)<1$. This has recently been theoretically shown in the context of optomechanics~\cite{Lemonde2014}. Using these models we numerically calculate $g^{(2)}(0)$ for a general mechanical single-mode Gaussian state undergoing squeezing.  For this, we use the most favorable parameters observed in our correlation experiments. We start from an initial thermal state $\hat{\rho}_{init}$ with $n_{init}=0.20$ phonons, which corresponds to the lowest temperature observed in the correlation measurements. The state is assumed to be purely thermal, which is in agreement with the experimentally observed autocorrelation of the pump pulse of $g^{(2)}(0) = 2.0^{+0.1}_{-0.1}$. Neglecting any heating from the optical pulses, we apply displacement $\hat{D}(\alpha)=\exp[\alpha\hat{b}^\dag-\alpha^*\hat{b}]$ and squeezing operations $\hat{S}(\xi)=\exp[\frac{1}{2} (\xi^{*}\hat{b}^2-\xi\hat{b}^{\dag 2})]$ with variable $\alpha = \bar{\alpha}\mathrm{e}^{i\phi}$ and $\xi = r\mathrm{e}^{i\theta}$ ($\bar{\alpha}, r > 0$). We then numerically minimize $g^{(2)}(0)$ of the resulting states $\hat{\rho} = \hat{D}(\alpha)\hat{S}(\xi)\hat{\rho}_{init}\hat{S}^\dag(\xi)\hat{D}^\dag(\alpha)$ as a function of $\alpha$ and $\xi$ using the quantum toolbox QuTiP~\cite{Johansson2012,Johansson2013}. The minimal correlation we can obtain is $g^{(2)}(0)\approx 0.95$, with $r=0.44$ and $\bar{\alpha}=2.00$ for $\theta = 2\phi$, clearly exceeding our experimentally measured value, as shown in Fig.~\ref{SM:Fig:3}. This simple model therefore allows us to exclude, with a p-value of 0.002, the possibility that the states we generate are in fact squeezed Gaussian states.  When additionally limiting the mean occupation $n = \mathrm{Tr}[\hat{\rho}\hat{b}^\dag\hat{b}]$ to the experimentally observed occupation number of the heralded states, $1.25<n<1.90$, we get a minimum $g^{(2)}(0)\approx0.99$, rejecting the hypothesis of observing a squeezed state with even stronger statistical confidence.

\end{document}